\documentclass{aa}

\usepackage{txfonts}
\usepackage{graphicx}
\usepackage{natbib}
\bibpunct{(}{)}{;}{a}{}{,}
\newcommand{\bv}{B\!-\!V}

\begin{document}

\title{Stellar Evolution and Variability in the Pre-ZAHB Phase}

\author{V. Silva Aguirre\inst{1,2} \and M. Catelan\inst{1} \and A. Weiss\inst{2} \and A. A. R. Valcarce\inst{1}}

\institute{Pontificia Universidad Cat\'olica de Chile, Departamento de Astronom\'ia y Astrof\'isica, Av. Vicu\~na Mackenna 4860, 782-0436 Macul, Santiago, Chile\\ email: vsilva,mcatelan,avalcarc@astro.puc.cl  \and Max Planck Institute for Astrophyics, Karl-Schwarzschild-Str. 1, 85748 Garching, Germany\\ email: weiss@mpa-garching.mpg.de}

\date{in press}   

\abstract{One of the most dramatic events in the life of a low-mass star is the He flash, which takes place at the tip of the red giant branch and is followed by a series of secondary flashes before the star settles on the zero-age horizontal branch (ZAHB). Yet, no stars have ever been positively identified in this key phase in the life of a low-mass star (hereafter the ``pre-ZAHB'' phase).} 
{In this paper, we investigate the possibility that at least some pre-ZAHB stars may cross the instability strip, thus becoming variable stars whose properties might eventually lead to positive identifications. In particular, it has been suggested that some of the RR Lyrae stars with high period change rates ($\dot{P}$) may in fact be pre-ZAHB stars. Here we present the first theoretical effort devoted to interpreting at least some of the high-$\dot{P}$ stars as pre-ZAHB pulsators.} 
{We constructed an extensive grid of evolutionary tracks using the Garching Stellar Evolution Code (GARSTEC) for a chemical composition appropriate to the case of the globular cluster M3 (NGC~5272), where a number of stars with high $\dot{P}$ values are found. We follow each star's pre-ZAHB evolution in detail, and compute the periods and period change rates for the stars lying inside the instability strip, also producing pre-ZAHB Monte Carlo simulations that are appropriate for the case of M3.} 
{Our results indicate that one should expect of order 1 pre-ZAHB star for every 60 or so bona-fide HB stars in M3. Among the pre-ZAHB stars, approximately 22\% are expected to fall within the boundaries of the instability strip, presenting RR Lyrae-like pulsations. On average, these pre-ZAHB pulsators are expected to have longer periods than the bona-fide HB pulsators, and 76\% of them are predicted to show negative $\dot{P}$ values. While the most likely $\dot{P}$ value for the pre-ZAHB variables is $\approx -0.3$~d/Myr, more extreme $\dot{P}$ values are also possible: 38\% of the variables are predicted to have $\dot{P} < -0.8$~d/Myr.} 
{It appears likely, therefore, that some~-- but certainly not all~-- of the RR Lyrae stars in M3 with high (absolute) $\dot{P}$ values are in fact pre-ZAHB pulsators in disguise.}


\maketitle

\section{Introduction}\label{intro}
Low-mass stars, after exhausting their central hydrogen supply, develop increasingly more massive and more highly degenerate helium cores as they continue their evolution accross the Hertzsprung-Russell diagram (HRD). As the mass of the core increases due to the continued supply of He-rich material from the H-burning shell that surrounds the core, it also contracts and becomes increasingly denser and hotter. Eventually, the conditions in the core become ripe for the ignition of the He-burning triple-$\alpha$ reactions. Since the core at this point is electron-degenerate, the process does not take place quiescently, but rather in the form of a thermonuclear runaway~-- the so-called {\rm He flash}. The primary He flash, which does not take place at the center proper (due to the very efficient neutrino cooling there), is followed by a series of secondary flashes, each occurring increasingly closer to the star's center, until degeneracy is lifted throughout the core and quiescent He burning finally commences~-- and a so-called {\em horizontal branch} (HB) star is born \citep[see][for a recent review]{mc07}.   

In the process just described, a low-mass star will typically change its luminosity by several orders of magnitude; its temperature may also increase very significantly. The inner structure of the star undergoes dramatic changes in the process as well. And yet, in spite of the importance of this evolutionary phase~-- hereafter the pre-zero-age HB (or pre-ZAHB) phase, to the best of our knowledge not a single such star has ever been positively identified. This dramatically limits the possibility of directly testing current models of pre-ZAHB evolution. 

The reasons why it has proven so challenging to pinpoint pre-ZAHB stars are twofold: first, the pre-ZAHB phase is very short, compared to other major evolutionary phases in the life of a low-mass star. Second, pre-ZAHB stars are largely expected to overlap the loci occupied by asymptotic giant branch (AGB), HB, and red giant branch (RGB) stars in observed color-magnitude diagrams (CMD).   

Bearing these difficulties in mind, \citet{mc05,mc07} suggested an alternative approach for the detection of pre-ZAHB stars, namely, through {\em stellar pulsations}. Indeed, some pre-ZAHB stars are expected to rapidly cross the Cepheid/RR Lyrae instability strip on their route from the RGB tip to the ZAHB, thus becoming pulsating stars along the way. At least some of these variables may present anomalously large period change rates ($\dot{P}$), as a consequence of their very high evolutionary speeds, and so at least some pre-ZAHB stars could be singled out in view of their high $\dot{P}$ values. As well known, high-$\dot{P}$ values (e.g., $|\dot{P}| \gtrsim 0.1-0.15$~d/Myr) have indeed often been observed among field and cluster RR Lyrae stars \citep[see, e.g.,][for a review]{hs95}, whereas canonical stellar evolution theory does not predict such high $\dot{P}$ values for these stars, except towards the end of the HB phase when the star is already approaching the AGB \citep[e.g.,][]{ywl91}. On the other hand, small, random mixing events associated with the composition redistribution in the cores of HB stars have also been proposed as possible causes for such high period changes in RR Lyrae stars \citep{sr79}.

The main goal of the present paper is, accordingly, to perform the first systematic study of the expected properties of pre-ZAHB stars, with a view towards their detection using a combination of pulsation and CMD properties. In this paper we shall, in particular, attempt to describe the expected pre-ZAHB stellar evolution in the case of the Galactic globular cluster M3 (NGC~5272), which is well known for being a particularly RR Lyrae-rich cluster, and also for containing a number of high-$\dot{P}$ RR Lyrae stars \citep[see, e.g., the recent work by][and references therein]{cc01}. 

In \S\ref{metodos} we describe the method employed to compute the evolutionary tracks, along with our Monte Carlo technique to produce synthetic pre-ZAHB distributions. In \S\ref{empirico} we present the empirical data that we use to compare our theoretical predictions with the observations. In \S\ref{resultados} we show the results from this comparison. Conclusions and final remarks are provided in \S\ref{conclusiones}.

\section{Methodology}\label{metodos}
\subsection{Evolutionary Tracks and Interpolation Grid}\label{tracks}
We used the Garching Stellar Evolution Code (GARSTEC) to construct the evolutionary tracks required to populate the pre-ZAHB phase. GARSTEC is a one-dimensional, hydrostatic code, which does not include the effects of rotation. The program in its present version is capable of calculating precise solar models as well as following the evolution of low-mass stars into the latest phases of their evolution, and has been successfully used to follow a star through the He flash \citep[e.g.,][]{hsea01}. 

In order to compute the present set of evolutionary tracks, we adopted the mixing length theory of convection with a value of $\alpha_\mathrm{\rm MLT}=1.75$, as obtained from a calibration of the solar model using the \citet{gs98} mixture. A chemical composition of $X=0.749$ (where $X$ is the hydrogen abundance) and $Z=0.001$ was used, as appropriate for M3. We used the OPAL equation of state \citep{frea96}, Alexander's opacities for low temperatures \citep{af94}, and OPAL opacities for high temperatures \citep{ir96}. The observed enhancement in the $\alpha$-capture elements among stars in the cluster \citep{cjea05} was taken into account in the computation of our opactity tables. The interested reader is referred to \citet{sw05} and \citet{ws00,ws07} for a much more detailed description of the numerics and input physics that go into this code.

Our models were computed with an initial mass on the zero-age main sequence (ZAMS) of $0.85 \, M_\odot$. This value was chosen in order to accomodate the current estimated age of the Universe, i.e. $13.7 \pm 0.2$~Gyr \citep{dsea03}, considering that, according to our computations, a $0.80 \, M_\odot$, non-mass-losing star reaches the ZAHB at an age of $\sim13$~Gyr. Mass loss was applied in its most widely used formulation, namely the \citet{dr75,dr77} formula, with a range of values for the corresponding mass loss ``efficiency parameter'' $\eta_\mathrm{R}$ that allowed us to produce from very red to very blue HB stars. While it is now known that the Reimers formulation does not properly describe mass loss in red giants \citep[e.g.,][and references therein]{sc05,mc08}, the specific choice of mass loss formalism turns out to be unimportant for our present purposes.  

\begin{figure*}
\centering
\resizebox{\hsize}{!}{\includegraphics{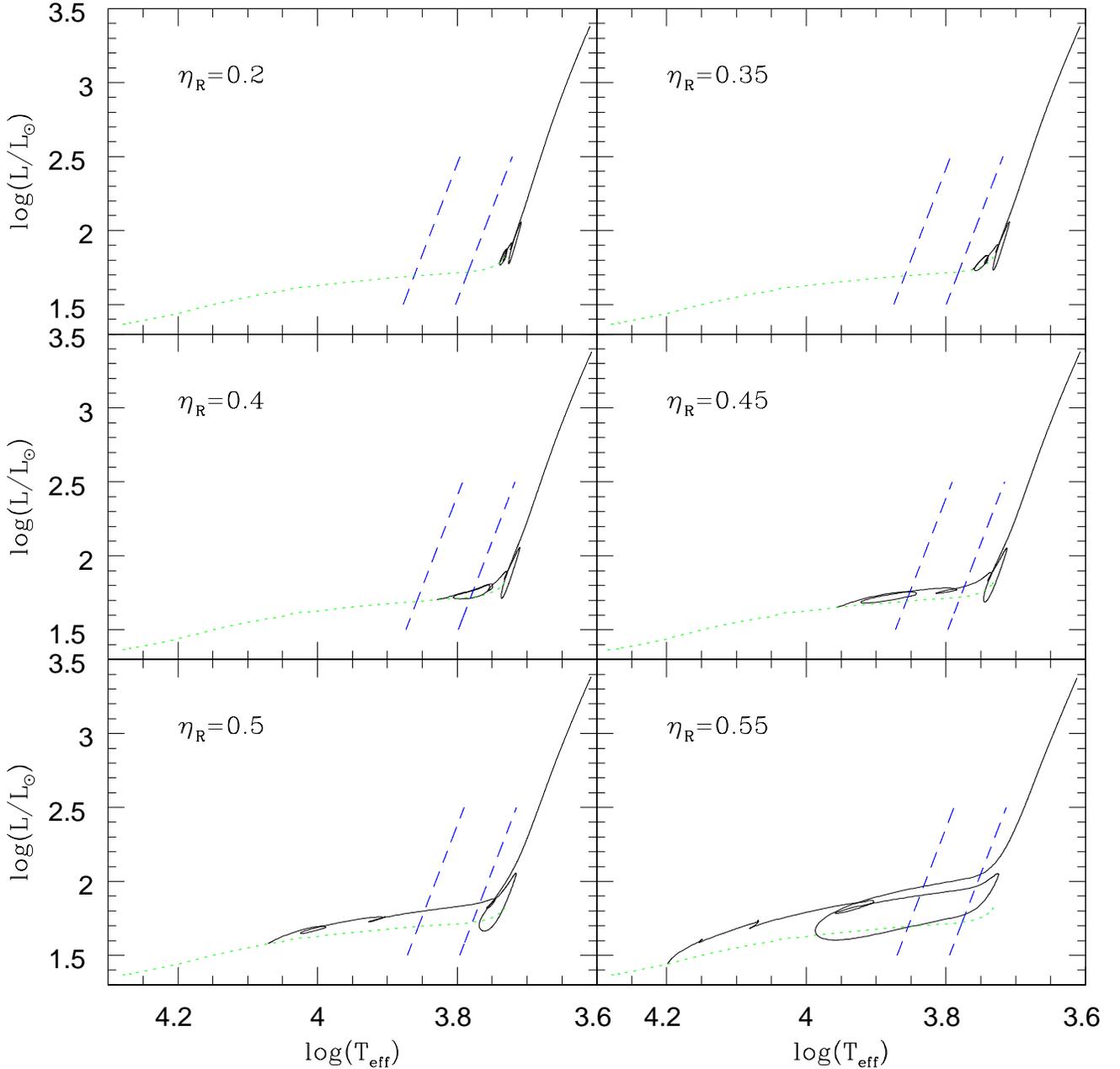}}
\caption{Pre-ZAHB evolutionary tracks ({\em solid lines}), based on full evolutionary calculations (extending from the ZAMS to the ZAHB) using different values of the Reimers mass loss parameter $\eta_\mathrm{R}$ (indicated on the upper left side of each panel). The edges of the instability strip are also schematically presented ({\em blue long-dashed lines}), whereas the theoretically derived ZAHB is depicted as a {\em green short-dashed line}.}
\label{calc_res}
\end{figure*}

A thin interpolation grid was created by evolving stars from the ZAMS to the ZAHB for a range of $\eta_\mathrm{R}$ values, from $\eta_\mathrm{R}=0$ to $\eta_\mathrm{R}=0.6$, in steps of 0.01. Following this procedure, a total of 61 evolutionary tracks was computed. As an example of the results of our calculations, we show pre-ZAHB evolutionary tracks for different $\eta_\mathrm{R}$ values in the HRD of Figure~\ref{calc_res}, where one can clearly see the secondary loops associated with each successive secondary flash (see also \S\ref{intro}).

According to our calculations, the total duration of the pre-ZAHB phase, from the RGB tip to the ZAHB, ranges from $\sim1.12$~Myr to $\sim1.4$~Myr, depending on the total mass of the star by the time it reaches the RGB tip. The criteria adopted to define the ZAHB location are the same as proposed by \citet[][and private communication]{as94}, which we now proceed to describe. As well known, the main He flash is produced near the center, but not in the exact center of the star. The He-burning region in the core moves inward through a series of low-amplitude secondary flashes until it reaches the center. Associated with each of these secondary flashes is a temporary convective zone. This convection finally reaches the center of the star, but the core at that time is still not fully in thermal equilibrium. As a result, the outer edge of the convective core moves inward until thermal equilibrium is largely restored. A minimum in the convective core size is reached and the convective core starts to grow as expected during the HB evolution. In our calculations, we took the minimum in the convective core size just after convection reaches the center as indicating that the star has finally reached the ZAHB.

We emphasize that hydrodynamical effects are ignored in our computations. That this is a sufficiently good approximation is supported by the hydrodynamical calculations by \citet{rd96} and \citet*{ddea06}, which indeed show that the He flash is mainly a non-hydrodynamical event in the life of a low-mass star \citep[but see also][ for a different viewpoint]{mmea08}. We also note that, due to the fact that the He flash originally occurs off-center, a molecular weight inversion takes place that may lead to a Rayleigh-Taylor instability, and thus trigger thermohaline mixing. We also ignore such a process in this paper, since it has been shown \citep*[][see their \S{7b}]{rkea80} that the timescale for such mixing to reach the center of a pre-ZAHB star is of order $10^8-10^9$~yr, which is much longer than the pre-ZAHB lifetime ($\sim 10^6$~yr). We have repeated the \citeauthor{rkea80} calculations in the case of our present models, and found, on the basis of their eq.~(A7), a timescale that is longer by about an order of magnitude than originally estimated by those authors. This confirms that thermohaline mixing is not important during pre-ZAHB evolution.

\subsection{Population Synthesis Code}\label{popcode}
To produce synthetic pre-ZAHB distributions, a suitable mass distribution must be adopted. Here a Gaussian-shaped mass deviate at the tip of the RGB was assumed, following the results by \citet{vc07} that indicate that the mass distribution along the M3 ZAHB does not differ markedly from a normal shape. Small corrections were applied to the mean and standard deviation values reported by those authors, due to the different adopted initial He content in the two studies. More specifically, in their calculations they used the evolutionary tracks created by \citet{mcea98} with a He content of $Y=0.23$, while our tracks were evolved using $Y=0.25$. This $0.02$ difference makes our ZAHB bluer and brighter than theirs for a given ZAHB mass, resulting in slightly higher mass values at a given ZAHB temperature. This leads to a normal deviate with $\left\langle M\right\rangle=0.647 \, M_\odot$ and $\sigma=0.022 \, M_\odot$, as adopted in the present study. Our evolutionary track with $\eta_\mathrm{R}=0.41$ reaches the tip of the RGB with a mass value of $0.645 \, M_\odot$, very similar to the mean value of the Gaussian. This track will accordingly be used in several of the examples to follow. 

Masses for the synthetic stars populating the pre-ZAHB were generated randomly following the method described in \citet{gbme58}. The total pre-ZAHB lifetime for the star's mass is then computed by interpolating over the evolutionary tracks. A random number between 0 and 1 is then generated for each star, with 0 corresponding to the beginning of the pre-ZAHB phase (i.e., to the main He flash) and 1 to its end (i.e., to the ZAHB) {\em for the track in our set whose pre-ZAHB evolution takes longest}. The star's evolutionary time (in Myr since the main He flash) is then computed by multiplying this random number by the maximum pre-ZAHB lifetime in our set of evolutionary tracks. (Underlying this procedure is the assumption that, apart from random fluctuations, stars are fed into the pre-ZAHB phase at an essentially constant pace.) Such an evolutionary time is then compared against the total pre-ZAHB lifetime {\em for the star's mass}, as obtained previously. As a consequence, some stars might have an assigned age that is {\em older} than their maximum calculated pre-ZAHB lifetime; these stars are discarded, since they are actually in a more evolved stage (i.e., the HB proper) than the pre-ZAHB phase. Once the mass of the synthetic pre-ZAHB star and its age have thus been assigned, the program obtains its luminosity and temperature by means of interpolations over our grid of evolutionary tracks. We are then in a position to check the star's variability status, and~-- if applicable~-- to compute its pulsation period. 

As can be seen from Figure~\ref{calc_res}, pre-ZAHB stars are expected to mostly occupy a locus similar to that of HB stars in the HRD. In addition, except for their innermost regions, the structure of these stars should be very similar to those of bona-fide HB stars occupying a similar position in the HRD. Therefore, any pulsating pre-ZAHB stars should be very similar to bona-fide RR Lyrae stars, and we can accordingly use theoretical predictions for the instability strip edges and pulsation periods of RR Lyrae stars to predict the pulsation properties of pre-ZAHB stars. 

In this sense, to define the blue edge of the instability strip, we use equation~(1) in 
\citet{fcea87}, but applying a shift by $-200$~K to the temperature values thus 
derived. The width of the instability strip has been taken as 
$\Delta\log T_{\rm eff}=0.075$; this provides the temperature of the 
red edge of the instability strip for each star once its blue edge has been 
determined. These choices provide a good agreement with more recent theoretical 
prescriptions and the observations \citep[see \S6 in][ for a detailed discussion]{mc04}. 

We include both fundamental-mode and first-overtone pre-ZAHB pulsators in our simulations. The computed periods for those stars are based on equation~(4) in 
\citet{fcea98}. Therefore, to compare our model prescriptions with the observations, the observed periods of any first-overtone pre-ZAHB variables must be ``fundamentalized,'' which can be achieved by adding 0.128 to the logarithm of the period \citep[e.g.,][and references therein]{mc05}. 

Period change rates are then computed for each synthetic star by following in detail the variation in $P$ with time along the interpolated evolutionary track for the star's mass. 

Let us focus on a synthetic star with mass $M=0.648 \, M_\odot$, and age since the He flash randomly assigned as 1.119~Myr. By interpolation on our grid, we find for this synthetic star a position in the HRD characterized by $\log(L/L_\odot)=1.711$, $\log(T_{\mathrm{eff}})=3.827$. This places the star inside the instability strip, pulsating with a period $P=0.546$~days~-- the latter decreasing at a rate $\dot{P}=-0.389$~d/Myr. 

\begin{figure}
\resizebox{\hsize}{!}{\includegraphics{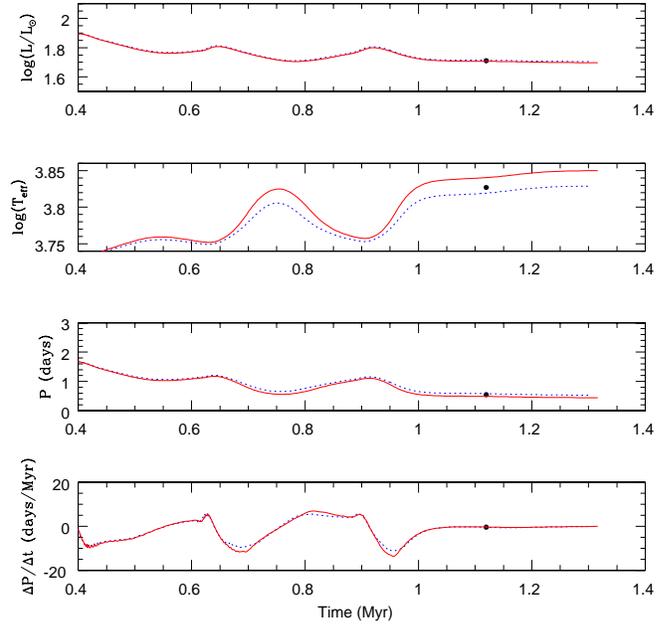}}
\caption{Time evolution for the $\eta_\mathrm{R}=0.41$ ({\em red, solid}) and $\eta_\mathrm{R}=0.40$ ({\em blue, dashed}) tracks. The {\em black dot} represents the position of one of the stars, as generated by our interpolation code, whose mass places it between these two evolutionary tracks. \textit{Upper panel}: luminosity evolution. \textit{Second panel}: temperature evolution. \textit{Third panel}: period evolution. \textit{Lower panel}: period change rate evolution. See text for details.}
\label{dtest}
\end{figure}

The result of the interpolation procedure is shown in Figure~\ref{dtest}. According to the star's mass value, it should be positioned between our evolutionary tracks for $\eta_\mathrm{R}=0.40$ and 0.41. The figure shows the time evolution of the physical parameters along these tracks (blue for $\eta_\mathrm{R}=0.40$ and red for $\eta_\mathrm{R}=0.41$). The time range has been selected to emphasize the time when the tracks cross the instability strip (i.e., $3.75 \leq \log T_{\mathrm{eff}} \leq 3.85$, approximately). As can clearly be seen, the star's position (indicated as a black dot) has been correctly interpolated between the two tracks. From the third panel, it can also be appreciated that changes in the trend of increasing or decreasing period values are indeed accompanied by zero values for $\dot{P}$ (fourth panel), and also by changes in the slope of the temperature curve (second panel). These changes correspond to the secondary flashes, which produce the previously noted wide loops along the evolutionary tracks in the HRD. It is also interesting to note that, at least in principle, very high (negative) period change rates (e.g., of order $-15$~days/Myr) are possible, although not very likely (see Fig.~\ref{histo_pdot})~-- and indeed, no stars with such extreme $\dot{P}$ values have been detected in M3. We will come back to this point later in this paper.

\begin{figure}
\resizebox{\hsize}{!}{\includegraphics{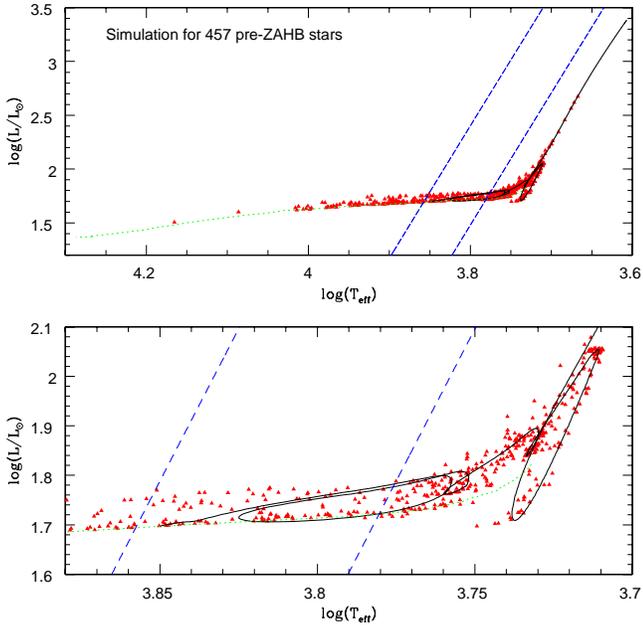}}
\caption{Results for one simulation with 457 pre-ZAHB stars. {\em Red triangles} indicate the synthetic pre-ZAHB stars. \textit{Upper Panel}: The {\em solid line} depicts the evolutionary track for a value of the Reimers mass loss parameter of $\eta_\mathrm{R}=0.41$, whereas the {\em green dotted line} shows the position of the ZAHB for different mass values. \textit{Lower Panel}: As in the upper panel, but zooming in around the secondary loops of the evolutionary track, and including the instability strip boundaries ({\em blue dashed lines}).}
\label{sim_res}
\end{figure}

\section{Empirical Data}\label{empirico}
To effectively compare our theoretical predictions with the observations, we need a database containing both variable and non-variable stars. Magnitudes and colors for the variable stars in \object{M3} were obtained from \citet{ccea05}, who provided quantities for the ``equivalent static star'' for a total of 133 variable stars in the cluster. For the non-variable stars, we used the photometry provided by \citet{rbea94} and \citet{ffea97}.

Our evolutionary tracks were converted from the theoretical [$\log(L/L_\odot)$,\,$\log(T_\mathrm{eff})$] to the observational [$M_V$,\,$(\bv)$] plane using the color transformations and bolometric corrections provided by \citet{dv03} for a metallicity  $[{\rm Fe/H}]=-1.5$ and abundance of the alpha elements $[\alpha/{\rm Fe}]=+0.3$. Interpolations on their tables was carried out using the algorithm by \citet{gh82}. We converted the thus derived magnitudes to apparent magnitudes using a distance modulus of 15.14 mag in $V$, based on a comparison between our evolutionary tracks and the observational data. This is only slightly different from the value provided by \citet{wh96}, namely 15.12~mag. Magnitudes and colors were corrected for extinction based on a reddening value $E(\bv) = 0.010$ \citep{wh96}, and adopting a standard extinction law with $A_V/E(\bv) = 3.1$.

\section{Results}\label{resultados}
\object{M3} has $\sim 530$ HB stars whose positions on the CMD have been measured \citep[][and references therein]{mc04}. As shown in Figure~\ref{sim_res}, the pre-ZAHB variable stars are expected to be primarily located above the ZAHB, and thus at least partially superimposed on the locus occupied by the bona-fide HB stars \citep[compare, e.g., with the CMD presented by][]{cc01}.

If we consider lifetimes, pre-ZAHB evolutions lasts, for the range of masses we are interested in, $\simeq 1.3$~Myr. We evolved our $\eta_\mathrm{R} = 0.41$ star up to the end of the HB phase, where He is exhausted in the core, and the evolution took $\simeq 77$~Myr. Consequently, one should expect to find one pre-ZAHB star for every 60 or so bona-fide HB stars. Therefore, among the 530 HB stars in M3, approximately 9 should be in the pre-ZAHB phase~-- two of which being pulsating (RR Lyrae-like) stars. Do the latter's expected pulsation properties differ in any noteworthy way from those characterizing bona-fide RR Lyrae stars?

\subsection{Expected Pulsation Properties of Pre-ZAHB Stars}
To answer this question, we ran pre-ZAHB Monte Carlo simulations for an M3-like input mass distribution. We used a total of 600 stars, of which $\sim 450$ are still in the pre-ZAHB phase, the remaining having already evolved past the ZAHB. (Since the age of the star is generated randomly, some of them do fall in the pre-ZAHB phase for a given mass value, while others have already left this evolutionary stage.) The latter are discarded, since we are only interested in the evolution prior to the ZAHB. The output from one of these simulations is shown in the HRD of Figure~\ref{sim_res}, where the position of the instability strip is also shown, along with the pre-ZAHB track for $\eta_\mathrm{R}= 0.41$, as well as the ZAHB locus obtained from our tracks for different mass values.

As can be seen from this plot, there is a considerable number of pre-ZAHB stars lying within the boundaries of the instability strip, and which can accordingly be tagged as variable stars. In this simulation approximately 100 variable stars were produced, corresponding to 22\% of the total number of pre-ZAHB stars in the simulation. Therefore, in the case of a globular cluster like M3, we expect that approximately one in every four or five pre-ZAHB stars will be disguised as bona-fide RR Lyrae stars.

\begin{figure}
\resizebox{\hsize}{!}{\includegraphics{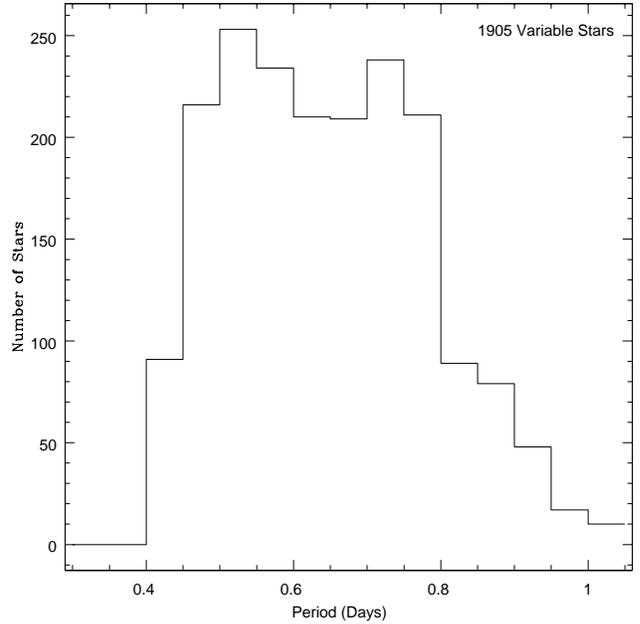}}
\caption{Distribution of fundamentalized periods from our simulations, for the 1905 pre-ZAHB variable stars obtained from our simulation for a total of 9427 pre-ZAHB stars}
\label{histo_per}
\end{figure}

\begin{figure}
\resizebox{\hsize}{!}{\includegraphics{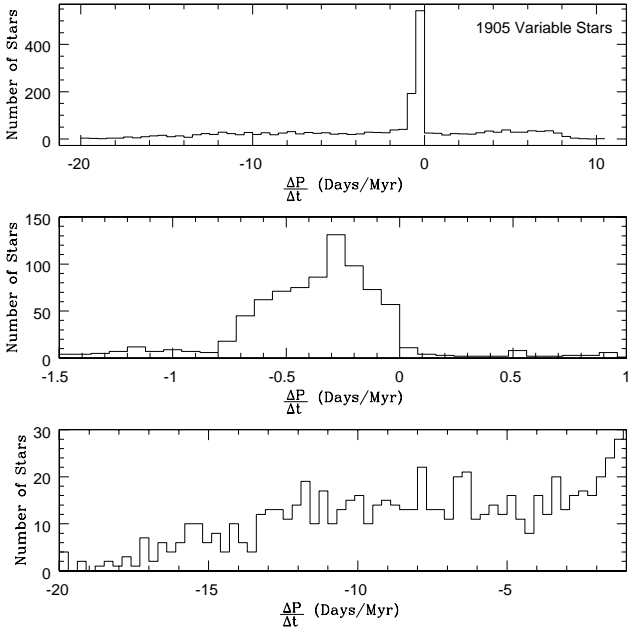}}
\caption{As in Figure~\ref{histo_per}, but for the period change rate distribution. The {\em middle panel} presents the same data as the upper panel, but zooming in around the peak of the distribution, whereas the {\em lower panel} shows the wing of the distribution for the more extreme (negative) $\dot{P}$ values.}
\label{histo_pdot}
\end{figure}

In Figures~\ref{histo_per} and \ref{histo_pdot} we present histograms that show the expected period and period change rate distributions for an M3-like input mass distribution. To minimize the effects of statistical fluctuations, we increased the number of stars in the Monte Carlo simulations up to a total of 10000 stars, 9427 of which turned out to be on the pre-ZAHB phase. The latter are the stars that are plotted in these figures. 

It is worth noting that, according to the derived period histogram, the expected periods of pre-ZAHB pulsators are relatively long, especially in view of the fact that \object{M3} is the prototype of an \citet{po39} type I cluster, where the average periods of the fundamental-type RR Lyrae are of order 0.55~days \citep[see, e.g.,][]{ccea01}. Indeed, the peak of the fundamentalized period distribution is found at 0.55~d in M3, and there are only a few RR Lyrae stars in the cluster with periods longer than 0.65~d \citep[see the corresponding histogram in][]{mc04}.

\begin{table*}
\centering
\caption{Average Colors, Magnitudes, and Periods as a Function of $\dot{P}$}
\label{tab:means}
\begin{tabular}{ccccccc}
\hline\hline
Range (d/Myr) &  $\langle V \rangle$ &  $\sigma_{\langle V \rangle}$  & $\langle \bv \rangle$ & $\sigma_{\bv}$  &   $\langle P \rangle$(day)  &  $\sigma_{P}$(day) \\
\hline
$\dot{P} < -0.8$            &  15.56  &   0.054   &  0.343  &   0.062  &  0.70  &  0.14 \\
$-0.8 < \dot{P} < 0$        &  15.66  &   0.015   &  0.320  &   0.061  &  0.59  &  0.11 \\
$\dot{P} < 0$               &  15.61  &   0.061   &  0.331  &   0.063  &  0.64  &  0.14 \\
$\dot{P} > 0$               &  15.63  &   0.053   &  0.338  &   0.061  &  0.66  &  0.13 \\
\hline
\end{tabular}
\end{table*}

Period {\em changes} are related to changes in the structure of a star through the period-mean density relation. More specifically, the periods of variable stars whose temperatures are decreasing in time, or whose luminosities are increasing, are expected to also increase. These concepts successfully explain the average behavior observed in globular clusters, where objects with bluer HB types~-- where a higher number of redward-evolving RR Lyrae stars is expected~-- tend to have higher average $\dot{P}$ values than those with redder HB morphology \citep[see, e.g., \S9 in][]{mc05}. As can be seen from Figure~\ref{histo_pdot}, most pre-ZAHB stars have period change rates contained in the $[-1,0]$~d/Myr range, meaning that evolution occurs mainly towards higher temperatures and/or decreasing luminosity~-- as expected for a pre-ZAHB star (see also Fig.~\ref{calc_res}). 

\begin{figure}
\resizebox{\hsize}{!}{\includegraphics{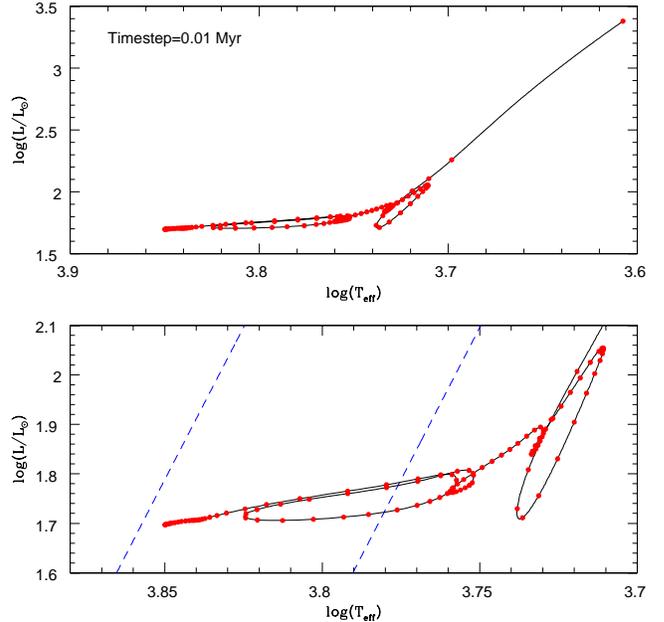}}
\caption{\textit{Upper Panel}: Pre-ZAHB evolution on the HRD for the star with $\eta_\mathrm{R}=0.41$. The evolutionary track is shown as a {\em solid line}, and {\em red dots} are plotted for every 0.01~Myr of evolution since the main flash. \textit{Lower Panel}: As above, but zooming in around the secondary loops, and including the instability strip edges ({\em dashed lines}).}
\label{time_res}
\end{figure}

To better understand this result, the time distribution along the evolutionary track corresponding closely to the most likely mass among M3 pre-ZAHB stars is presented in Figure~\ref{time_res}. In this figure, a red dot was placed for every 0.01~Myr of evolution, starting from the He flash and up to the ZAHB. As we have seen, the period changes along the blue loops can reach values up to $\sim -15$~d/Myr, which is due to the very fast evolution of stars in the secondary loops. However, this is a relatively unlikely situation (see the lower panel in Fig.~\ref{histo_pdot}), since pre-ZAHB stars spend most of their time in the blueward evolution that takes place after the last secondary flash, immediately before reaching the ZAHB. This implies that most (i.e., $\simeq 76\%$) pre-ZAHB variables will actually have {\em negative} $\dot{P}$ values~-- as confirmed by Figure~\ref{histo_pdot}, which in addition reveals that the most likely $\dot{P}$ value for a pre-ZAHB star is $\approx -0.3$~d/Myr. As a consequence, according to our simulations, only 24\% of the pre-ZAHB stars, for an M3-like HB morphology, are expected to present positive $\dot{P}$ values. On the other hand, 38\% of the pre-ZAHB pulsators present $\dot{P}$ values more negative than $-0.8$~d/Myr. We again emphasize that such extreme $\dot{P}$ values are completely unexpected on the basis of canonical HB evolution, according to which period change rates should not be higher (in absolute value) than about 0.1~d/Myr in a cluster such as M3 \citep[see, e.g., Fig.~2 in][]{ywl91}.

A closer look at our results is provided in Table~\ref{tab:means}, where we show the average $V$-band magnitudes and $\bv$ colors, along with the mean fundamentalized periods (and corresponding standard deviations), for different groups of stars with different $\dot{P}$ values in our M3 simulations. The most noteworthy trend present in this table is for the stars with $\dot{P} < -0.8$~d/Myr to be slightly brighter, and have correspondingly longer periods, than those with smaller $\dot{P}$ values. However, these trends are far from constituting firm tendencies, given the fairly large standard deviations that accompany the studied quantities.

\begin{figure}
\resizebox{\hsize}{!}{\includegraphics{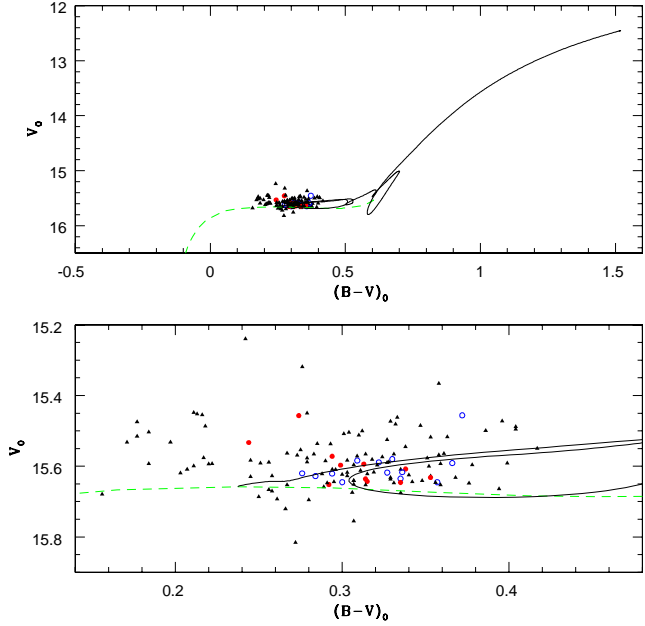}}
\caption{\textit{Upper Panel}: CMD for M3, including only the RR Lyrae variable stars. {\em Filled red circles}: stars with $\dot{P} < -0.15$~d/Myr; {\em open blue circles}: stars with $\dot{P} > +0.15$~d/Myr; {\em triangles}: RR Lyrae stars with smaller (or undetermined) $\dot{P}$ values. The evolutionary track ({\em solid line}) for an $\eta_\mathrm{R}=0.41$, as well as the ZAHB ({\em green dashed line}), are overplotted on the observed data. \textit{Lower Panel}: Same as in the previous panel, but zooming in around the instability strip.}
\label{teo_plan}
\end{figure}

\begin{figure}
\resizebox{\hsize}{!}{\includegraphics{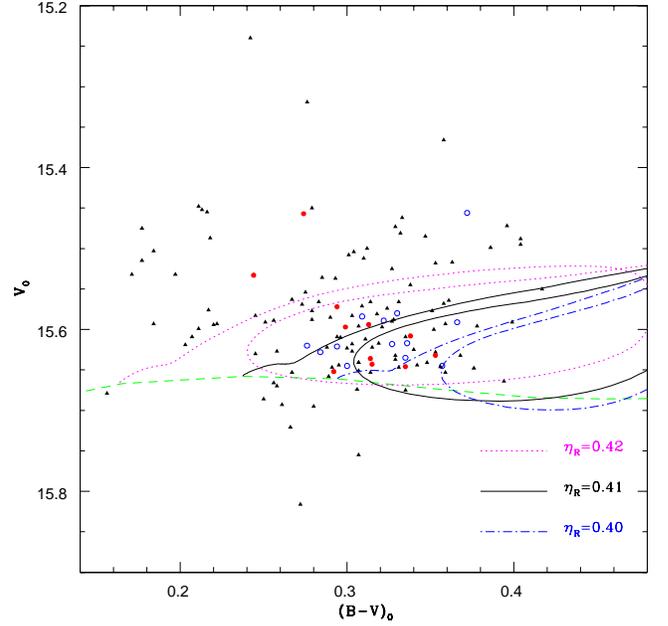}}
\caption{Distribution of M3 RR Lyrae stars on the CMD (symbols are the same as in Fig.~\ref{teo_plan}), with three evolutionary tracks superimposed: \textit{dashed magenta line}: track for $\eta_\mathrm{R}=0.42$; \textit{solid line}:
 track for $\eta_\mathrm{R}=0.41$; \textit{dash-dotted blue line}: track for $\eta_\mathrm{R}=0.40$. Masses at the tip of the RGB are $0.639\, M_\odot$, $0.645\, M_\odot$, and $0.650\, M_\odot$, respectively. See text for mode details.}
\label{teo_plan2}
\end{figure}

In Figure~\ref{teo_plan}, we present a plot showing the variable stars in \object{M3} \citep[data from][]{ccea05}, together with the ZAHB and the pre-ZAHB evolutionary track for $\eta_\mathrm{R}=0.41$. Stars with measured period change rates higher than $+0.15$~d/Myr are plotted as open circles, whereas those with $\dot{P} < -0.15$~d/Myr are shown as filled circles. In Figure~\ref{teo_plan2} we present the same variable stars but now including three evolutionary tracks: for $\eta_\mathrm{R}=0.42$, $\eta_\mathrm{R}=0.41$, and $\eta_\mathrm{R}=0.40$. These tracks were selected because their associated mass values at the tip of the RGB ($0.639\, M_\odot$, $0.645 \, M_\odot$, and $0.650 \, M_\odot$, respectively) make them representative of the majority of pre-ZAHB stars in M3, whose HB mass distribution is characterized by $\left\langle M\right\rangle=0.647 \, M_\odot$ and $\sigma=0.022 \, M_\odot$ (see \S\ref{popcode} for details). As can be seen from both these figures, the positions of these variables in the CMD are broadly consistent with at least some of them being pre-ZAHB stars~-- and indeed, according to our simulations, at least $\sim 2$ of them very likely are. 

\subsection{Variable vs. Non-Variable Pre-ZAHB Stars} 

As we have seen, our simulations indicate that of order 20\% of the pre-ZAHB stars are expected to be RR Lyrae-like variables. What about the remaining 80\% of these stars? 

To answer this question, we divide the pre-ZAHB stars into four different groups, depending on their photometric and pulsation characteristics, as follows: RR Lyrae-like, red HB-like, blue HB-like, and AGB/RGB like. The relative proportions of pre-ZAHB stars in each of these groups, according to our simulation with 9427 pre-ZAHB stars, are provided in Table~\ref{tab:taxonomia}. 

According to Table~\ref{tab:taxonomia}, most pre-ZAHB stars are expected to fall in the red HB region, with a proportion of almost three red HB-like pre-ZAHB stars for every RR Lyrae-like pre-ZAHB star. Comparable fractions of blue HB-like and RR Lyrae-like pre-ZAHB stars are also implied. But a minor fraction (1.4\%) is expected to be sufficiently red and bright (i.e., brighter than the expected terminal-age HB, or core helium exhaustion, locus) as to fall into the AGB/RGB-like category. This is a consequence of the extremely fast evolutionary rates immediately after the primary He flash (see Fig.~\ref{time_res}). In the case of M3, therefore, taking into account a total population of 530 HB stars, it appears most likely that there are $\sim 2$ RR Lyrae-like, $\sim 2$ blue HB-like, $\sim 5$ red HB-like, and an insignificant number~-- most likely 1 at most~-- AGB/RGB-like stars.

\begin{table}
\centering
\caption{Relative Proportions of Pre-ZAHB Stars in M3 Simulations with 9427 Pre-ZAHB Stars}
\label{tab:taxonomia}
\begin{tabular}{ccc}
\hline\hline
Classification &  Number &  Number fraction  \\
\hline
Red HB-Like              &  5407  &   57.4\%   \\
RR Lyrae-Like            &  1905  &   20.2\%   \\
Blue HB-Like             &  1979  &   21.0\%   \\
AGB/RGB-Like             &   136  &    1.4\%   \\
\hline
\end{tabular}
\end{table}

\section{Conclusions}\label{conclusiones}
In this paper, we have performed a detailed theoretical study of the expected evolutionary and pulsation properties of pre-ZAHB stars for a metallicity appropriate to the case of the Galactic globular cluster M3. Our results indicate that there should be around one pre-ZAHB star for every 60 or so bona-fide HB stars in the cluster. Based on extensive Monte Carlo simulations we find that, among the pre-ZAHB stars, of order 20\% should exhibit RR Lyrae-like pulsations~-- to be compared with 21\% and 57\% blue HB- and red HB-like pre-ZAHB stars, respectively, and but a very small (of order $1-2\%$) population of AGB/RGB-like pre-ZAHB stars. 

The pre-ZAHB, RR Lyrae-like pulsators are characterized by relatively long periods (typically longer than the average for the cluster), and (especially) by period change rates falling mostly in the $[-1,0]$~d/Myr range, with $-0.3$~d/Myr being the most likely value. However, more extreme $\dot{P}$ values are also possible; in fact, 38\% of the variables in our simulation have $\dot{P} < -0.8$~d/Myr. Relatively few (i.e., around 24\%) of the pre-ZAHB pulsators are found with positive period change rates. 

Our pre-ZAHB Monte Carlo simulations show, in addition, that for an M3-like HB mass distribution the pre-ZAHB CMD positions should be basically indistinguishable from those of bona-fide HB stars in the cluster. We predict that, in the specific case of M3, $\sim 9$ pre-ZAHB stars should be present (assuming a total population of 530 HB stars), $\sim 2$ of which most likely disguised as RR Lyrae stars with relatively long (and negative) period change rates. Since the number of stars in M3 with high $\dot{P}$ is significantly larger \citep[e.g.,][]{cc01}, we conclude that pre-ZAHB evolution may explain the high-$\dot{P}$ phenomenon for some (but not all) of the RR Lyrae stars with high period change rates in this cluster. 

In the future, we will extend our calculations to other clusters (and dwarf galaxies) with different metallicities and HB morphologies. These calculations should also prove important for the identification of candidate pre-ZAHB pulsators among field stars in the Galaxy and neighboring galaxies. With the era of the survey telescopes quickly approaching, the number of such stars available for study should dramatically increase, thus making empirical studies of pre-ZAHB evolution increasingly more feasible~-- at least in the long term.

\begin{acknowledgements}
The authors wish to thank Allen V. Sweigart for some helpful discussions. VS, MC and AV acknowledge support by Proyecto Fondecyt Regular \#1071002. The comments and suggestions by an anonymous referee are also gratefully appreciated. VS is partially supported by the Fondap Center for Astrophysics grant \#15010003. 
\end{acknowledgements}

\bibliographystyle{aa}

\end{document}